\documentclass[11pt]{article}
\usepackage{times}
\usepackage{sidecap}
\usepackage{geometry}
\usepackage{floatrow}
\usepackage{float}
\usepackage{hyperref}
\usepackage{graphics,graphicx}
\usepackage{subcaption}
\usepackage{epsf,wrapfig}
\usepackage{amsmath}
\usepackage{natbib,bm}
\usepackage{paralist}
\let\olditem\item
\renewenvironment{thebibliography}[1]{%
  \section*{\refname}
  \let\par\relax
  \renewcommand{\item}[1][]{\olditem[\textbullet]}%
  \inparaenum}{\endinparaenum}
\newcounter{aff}

\DeclareGraphicsExtensions{.jpg,.pdf,.png,.eps}

\usepackage{caption}
\usepackage{xcolor}
\usepackage{tcolorbox}
\bibpunct{(}{)}{;}{a}{}{,}

\usepackage{soul}
\definecolor{darkcyan}{rgb}{0.0, 0.0, 0.7}

\definecolor{boxcolor}{RGB}{193, 229, 255}
\definecolor{figboxcolor}{RGB}{193, 229, 255}
\definecolor{boxcolorgray}{RGB}{227, 227, 227}
\definecolor{darkcyan}{rgb}{0.0, 0.0, 0.7}

\sethlcolor{boxcolorgray}

\def\gtrsim{\lower 2pt \hbox{$\, \buildrel {\scriptstyle >}\over
{\scriptstyle \sim}\,$}}
\def\lesssim{\lower 2pt \hbox{$\, \buildrel {\scriptstyle <}\over
{\scriptstyle \sim}\,$}}

\def\approxlt{\lower.2em\hbox{$\buildrel < \over \sim$}}
\def\approxgt{\lower.2em\hbox{$\buildrel > \over \sim$}}

\def\be{\begin{equation}}
\def\ee{\end{equation}}
\def\bea{\begin{eqnarray*}}
\def\eea{\end{eqnarray*}}

\def\cii{[C\,{\sc ii}]}
\def\oi{[O\,{\sc i}]}

\def\oiii{[O\,{\sc iii}]}

\def\nii{[N\,{\sc ii}]}
\def\niii{[N\,{\sc iii}]}
\def\jref@jnl#1{{\rm#1}}
\def\aj{\jref@jnl{AJ}}			
\def\araa{\jref@jnl{ARA\&A}}		
\def\apj{\jref@jnl{ApJ}}			
\def\apjl{\jref@jnl{ApJ}}		
\def\apjs{\jref@jnl{ApJS}}		
\def\ao{\jref@jnl{Appl.~Opt.}}		
\def\apss{\jref@jnl{Ap\&SS}}		
\def\aap{\jref@jnl{A\&A}}		
\def\aapr{\jref@jnl{A\&A~Rev.}}		
\def\aaps{\jref@jnl{A\&AS}}		
\def\azh{\jref@jnl{AZh}}			
\def\baas{\jref@jnl{BAAS}}		
\def\jrasc{\jref@jnl{JRASC}}		
\def\memras{\jref@jnl{MmRAS}}		
\def\mnras{\jref@jnl{MNRAS}}		
\def\pra{\jref@jnl{Phys.~Rev.~A}}	
\def\prb{\jref@jnl{Phys.~Rev.~B}}	
\def\prc{\jref@jnl{Phys.~Rev.~C}}	
\def\prd{\jref@jnl{Phys.~Rev.~D}}	
\def\pre{\jref@jnl{Phys.~Rev.~E}}	
\def\prl{\jref@jnl{Phys.~Rev.~Lett.}}	
\def\pasp{\jref@jnl{PASP}}		
\def\pasj{\jref@jnl{PASJ}}		
\def\qjras{\jref@jnl{QJRAS}}		
\def\skytel{\jref@jnl{S\&T}}		
\def\solphys{\jref@jnl{Sol.~Phys.}}	
\def\sovast{\jref@jnl{Soviet~Ast.}}	
\def\ssr{\jref@jnl{Space~Sci.~Rev.}}	
\def\zap{\jref@jnl{ZAp}}			
\def\nat{\jref@jnl{Nature}}		
\def\iaucirc{\jref@jnl{IAU~Circ.}}
\def\aplett{\jref@jnl{Astrophys.~Lett.}}
\def\apspr{\jref@jnl{Astrophys.~Space~Phys.~Res.}}
\def\bain{\jref@jnl{Bull.~Astron.~Inst.~Netherlands}}
\def\fcp{\jref@jnl{Fund.~Cosmic~Phys.}}
\def\gca{\jref@jnl{Geochim.~Cosmochim.~Acta}}
\def\grl{\jref@jnl{Geophys.~Res.~Lett.}}
\def\jcp{\jref@jnl{J.~Chem.~Phys.}}	
\def\jgr{\jref@jnl{J.~Geophys.~Res.}}	
\def\jqsrt{\jref@jnl{J.~Quant.~Spec.~Radiat.~Transf.}}
\def\memsai{\jref@jnl{Mem.~Soc.~Astron.~Italiana}}
\def\nphysa{\jref@jnl{Nucl.~Phys.~A}}
\def\physrep{\jref@jnl{Phys.~Rep.}}
\def\physscr{\jref@jnl{Phys.~Scr}}
\def\planss{\jref@jnl{Planet.~Space~Sci.}}
\def\procspie{\jref@jnl{Proc.~SPIE}}

\def\ion#1#2{#1$\;${\small\rm\@Roman{#2}}\relax}

\newcount\lecurrentfam
\def\LaTeX{\lecurrentfam=\the\fam \leavevmode L\raise.42ex
\hbox{$\fam\lecurrentfam\scriptstyle\kern-.3em A$}\kern-.15em\TeX}
\def\sizrpt{
(\fontname\the\font): em=\the\fontdimen6\font, ex=\the\fontdimen5\font
\typeout{
(\fontname\the\font): em=\the\fontdimen6\font, ex=\the\fontdimen5\font
}}
\definecolor{dark}{rgb}{ 0.34, 0.09, 0.49}

\usepackage[utf8]{inputenc}
\usepackage[T1]{fontenc}
\usepackage{fancyhdr,url}
\usepackage{transparent}
\usepackage{tablefootnote}
\usepackage{xcolor}
\usepackage{mathpazo}

\begin{document}

\noindent
\begin{raggedright}
{\Huge\bfseries The dawn of the first galaxies \\
[0.5em]
\Large  Unveiling the interstellar medium of galaxies at $z>10$ with ALMA2040
}\end{raggedright}\\

\noindent
Livia Vallini$^{\dagger,1}$,
Tanio Diaz Santos$^{\dagger,2}$,
Belen Alcalde Pampliega$^{3}$,
Hiddo Algera$^{4}$,
Manuel Aravena$^{5}$,
Cecilia Bacchini$^{6,7}$,
Tom Bakx$^{8}$,
Maru\v{s}a Brada{\v c}$^{9}$,
Leindert A. Boogaard$^{12}$,
Karina I. Caputi$^{10}$,
Stefano Carniani$^{11}$,
Karin Cescon$^{12}$,
Francoise Combes$^{13}$,
Carla Cornil-Ba\"{i}otto$^{14}$,
Pierre Cox$^{15}$,
Elisabete Da Cunha$^{16}$,
Roberto Decarli$^{1}$,
Miroslava Dessauges-Zavadsky$^{17}$,
Andreas L. Faisst$^{18}$,
Roman Fernandez Aranda$^{19}$,
Andrea Ferrara$^{11}$,
Jonathan Freundlich$^{20}$,
Yoshinobu Fudamoto$^{21}$,
Seiji Fujimoto$^{23}$,
Carlotta Grup\-pio\-ni$^{1}$,
Takuya Hashimoto$^{24}$,
Natsuki Hayatsu$^{25}$,
Kasper E. Heintz$^{26,27,28}$,
Jacqueline Hodge$^{12}$,
Ryota Ikeda$^{12}$,
Gareth C. Jones$^{29,30}$,
Mahsa Kohandel$^{1}$,
Adarsh Kumar$^{31}$,
Alvaro Labiano$^{32}$,
Ivanna Langan$^{19}$,
Andrea Lapi$^{33}$,
Umberto Maio$^{34,35}$,
Marcella Massardi$^{36}$,
Dieu D. Nguyen$^{37}$,
Andrea Pallottini$^{38}$,
Mamta Pandey-Pommier$^{39}$,
Miguel Pereira-Santaella$^{40}$,
Gerg\H{o} Popping$^{7}$,
Francesca Pozzi$^{41,1}$,
Pra\-chi Prajapati$^{42,43}$,
Dominik A. Riechers$^{42,43}$,
Francesca Rizzo$^{10}$,
Lucie Rowland$^{12}$,
San\-der Schouws$^{44}$,
Irene Shivaei$^{19}$,
Laura Som\-mo\-vi\-go$^{45,46}$,
Yuma Sugahara$^{25}$,
Hannah R. Stacey$^{7}$,
Sandro Tacchella$^{29,30}$,
Eros Vanzella$^{1}$,
Enrico Veraldi$^{33}$,
Julie Wardlow$^{47}$,
Joris Witstok$^{27,28}$,
Anita Zanella$^{1}$

{\footnotesize{
\noindent{$\dagger$ These authors contributed equally to this work}\\
\noindent{\emph{Affiliations can be found after the references}}
}}
\begin{tcolorbox}[boxsep=3pt,left=0pt,right=0pt,top=1pt,bottom=1pt,colframe=boxcolor,colback=boxcolor, halign=justify]
\textbf{Abstract:} 
Understanding how the first galaxies assembled their stellar mass at cosmic dawn is a central frontier of modern astrophysics. JWST has now revealed an unexpected population of ultraviolet-bright galaxies at $z>10$, whose intense emission points towards a rapid stellar mass growth that challenges current models of early galaxy assembly. Yet, the cold gas fueling this growth remains largely hidden: current ALMA sensitivity restricts routine detection to only the brightest ionized-gas tracers such as \oiii{}\,88\,$\mu$m, while \cii{}\,158\,$\mu$m, a key coolant and gas-mass proxy of the neutral interstellar medium (ISM), remains undetected beyond $z\approx8$. This leaves fundamental questions unanswered: how quickly primeval galaxies gathered cold gas and turned it into stars; what physical conditions (density, temperature, radiation field) characterized their ISM; how fast metals and dust built up; and how feedback regulated early gas assembly. Only a transformational sub-millimeter facility like ALMA2040 can address these questions, enabling systematic surveys of \cii{} in the first galaxies and providing a representative census of gas fractions and depletion times. A simulated galaxy at $z=11.3$ shows that ALMA2040 not only will detect, but also spatially resolve its merging cold-gas structure in \cii{} on $\approx$\,200\,pc scales in 5\,hours (below 100\,pc in 10\,hours). ALMA2040 will also spectrally resolve multiple lines simultaneously, akin to integral-field spectroscopy, providing the key diagnostics needed to break density--metallicity--radiation-field degeneracies and map the multi-phase ISM of $z>10$ galaxies. Finally, the combination of dramatically increased line sensitivity and wide instantaneous bandwidth will push the search for the first galaxies out to the earliest cosmic times, with \oiii{} detectable up to $z\approx 20$. ALMA2040 will thus be transformative for understanding how the first galaxies assembled at cosmic dawn.

\end{tcolorbox}
\newpage{}

\section{Scientific context and open questions}
\begin{wrapfigure}[24]{r}{0.55\textwidth}
  \vspace{-5mm}
  \centering
  \includegraphics[width=\linewidth]{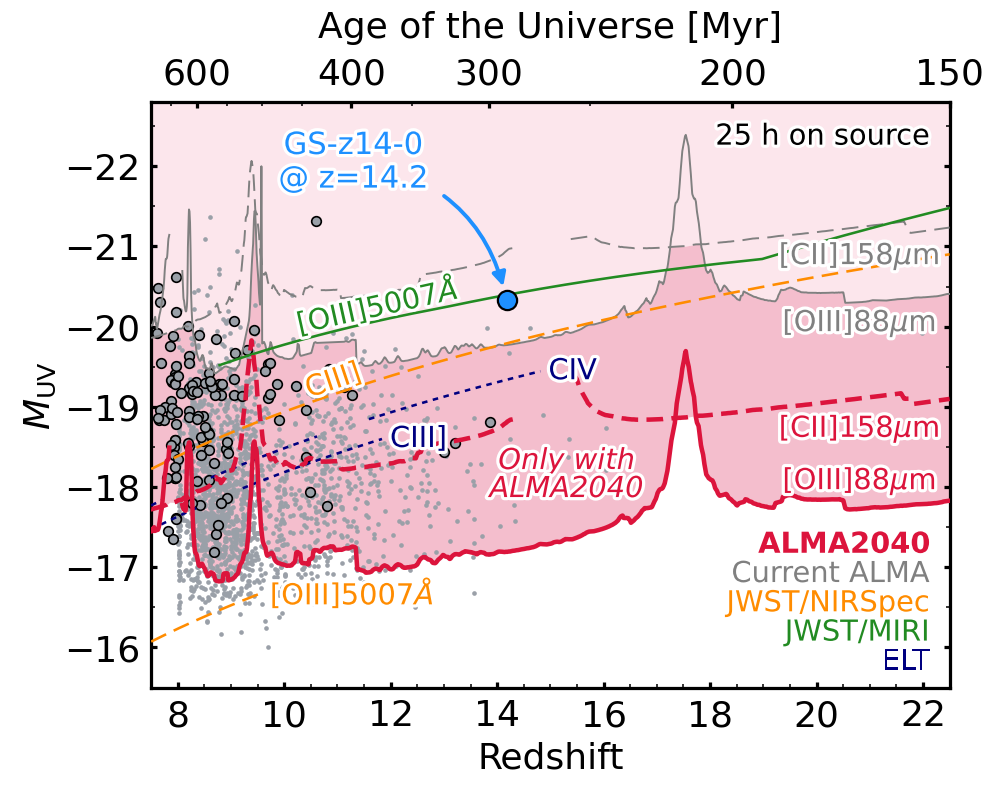}
  \caption{ALMA2040 can uniquely target emission lines from galaxies in the $z>8$ Universe. Black-edged points are sources with $z_{spec}$, while small grey points are those with $z_{phot}$ from JADES \citep[][]{Hainline2026}. Each curve marks the faintest UV magnitude at which a line is detected at 5\,$\sigma$ in 25\,hours on source using UV-\oiii{} and UV-\cii{} relations from \citet{Harikane2020} and \citet{Schaerer2020}. }
  \label{fig:muvplot}
  \vspace{-2mm}
\end{wrapfigure}

A major frontier in modern astrophysics is understanding how the first galaxies assembled and formed their stars during the Epoch of Reionization (EoR). Over the last decade, the advent of ALMA and, subsequently, JWST has revolutionized the field \citep[][]{Smit2026, Stark2026}, pushing the redshift frontier of the galaxies we can detect and characterize deep into the EoR. In particular, JWST, with its superb sensitivity and spatial resolution, is uncovering an unexpected abundance of luminous, and potentially massive galaxies at $z \gtrsim 10$ \citep[][]{Naidu2022, CurtisLake2023, Finkelstein2024, Carniani2025, Naidu2026}. However, current ALMA capabilities require prohibitively long integrations ($>$\,10\,h per target; see Fig.~\ref{fig:muvplot}) to resolve even a single line tracing the cold gas fueling star formation at cosmic dawn \citep{Schouws2025b}, leaving this critical reservoir far less accessible than the stellar and unobscured ionized gas traced by JWST. Moreover, although gravitational lensing can boost both spatial resolution and sensitivity \citep{DZ2019}, resolving sub-kiloparsec scales at $z>1$ remains constrained by ALMA's current baseline configurations. In other words, while JWST's thrilling discoveries are challenging our understanding of early galaxy formation, its capabilities are restricted to probing the ionized gas component, leaving our view of the ISM fundamentally incomplete. A sub-mm perspective on the cold gas phases is therefore crucial to address key open questions:

\begin{itemize}
  \setlength\itemsep{0pt}
  \setlength\parskip{0pt}
  \bf 
 \item  How fast do primeval galaxies gather their cold gas and form stars?
 \item  What are the interstellar medium conditions of typical galaxies at cosmic dawn?
 \item How fast was the metal enrichment and molecular build-up in the early Universe?
 \item What is the role of feedback on the formation of the first galaxies?
\end{itemize}

In what follows, we show that such a transformational facility is essential to answering these questions. By opening a direct window on the neutral and molecular gas fueling star formation, and on the ionized gas traced by rest-frame far-infrared (FIR) lines unaffected by dust extinction, ALMA2040 will overcome current limitations (Fig.~\ref{fig:muvplot}), securing spectroscopic redshifts and shedding light on how the first galaxies assembled and evolved at the dawn of cosmic history.

\subsection{How fast did primeval galaxies gather their cold gas and form stars?}
\label{s:content}
In the following two subsections we argue how ALMA2040 can be used to \textbf{both} push the redshift frontier, potentially identifying the farthest galaxies ever discovered, and to perform unprecedented spatially resolved analyses of the ISM of hundreds of galaxies beyond $z>10$.

\subsubsection{Detecting galaxies up to \texorpdfstring{$z\sim20$}{z=20} with ALMA2040}
\label{s:ultrahighz}

\begin{figure}[ht!]
    \centering
    \includegraphics[width=1.\linewidth]{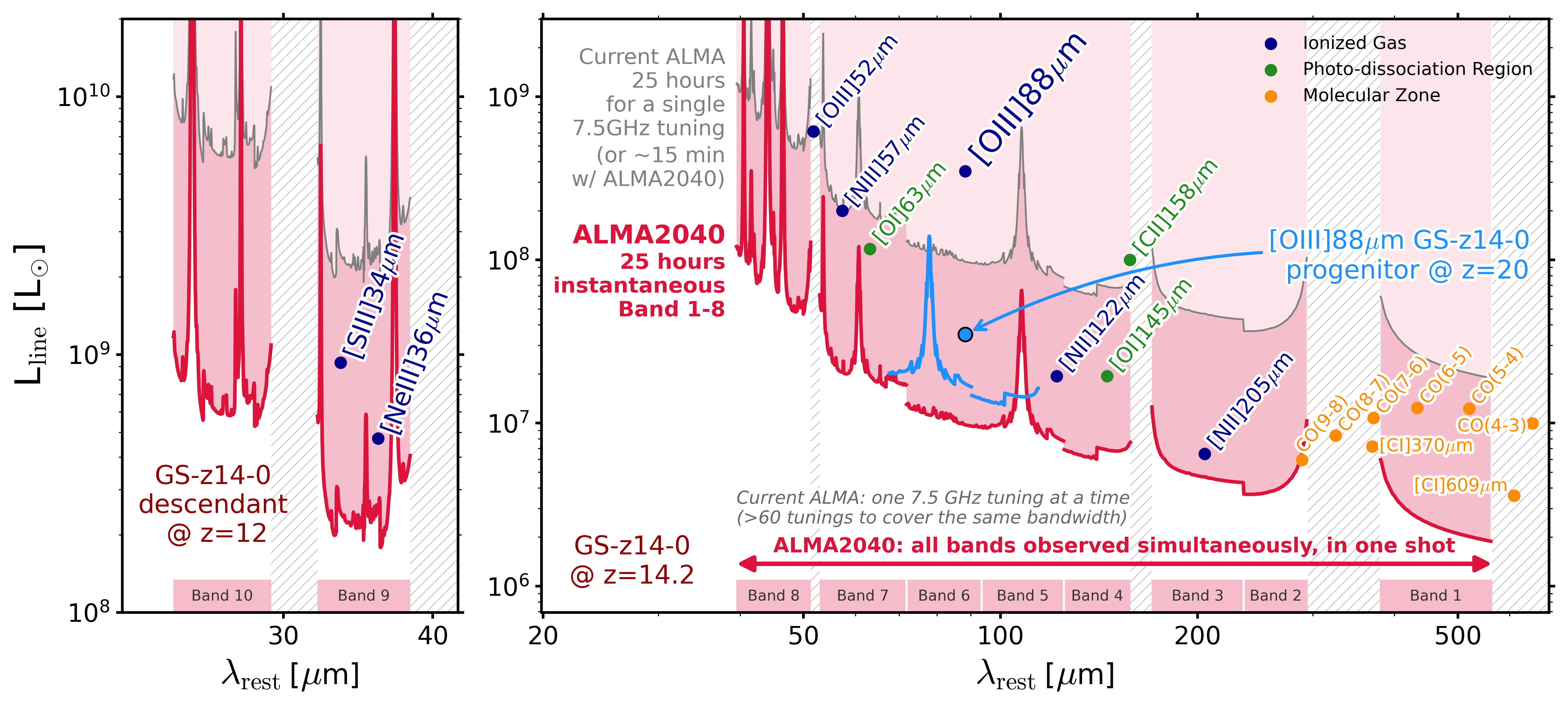}
    \caption{Point source line sensitivity (5\,$\sigma$, 25\,hours on source) of the current ALMA (gray line) and the projected ALMA2040 facility (red line), from Band 1 to 8 (right panel) and 9 to 10 (left panel). By increasing the number of antennas and improving receiver sensitivities, ALMA2040 will reach a depth $\geq$\,10$\times$ higher than the current ALMA; i.e., ALMA2040 will need $\geq$\,100$\times$ shorter integration times to reach the same sensitivity. As an example, observed and expected line luminosities for GS-z14-0, at $z=14.2$, are shown on top of the Bands\,1--10 sensitivities. It would take the current ALMA $\sim$\,25\,hours of on-source time to barely detect, \textit{\textbf{one-by-one}}, some of the most physics-constraining FIR lines, such as \oiii{}\,52\,$\mu$m, \niii{}\,57\,$\mu$m, \oi{}\,63\,$\mu$m or \cii{}\,158\,$\mu$m. Other critical ISM lines such as \nii{}\,122\,$\mu$m, \oi{}\,145\,$\mu$m or \nii{}\,205\,$\mu$m would remain unreachable. With ALMA2040 we will also be able to capture for the first time rest-frame MIR lines at the highest-frequencies from GS-z14-0-like descendants if they were, e.g., at $z=12$. Furthermore, ALMA2040 will obtain instantaneous, $\geq$\,512\,GHz bandwidth spectra, thus covering \textit{\textbf{all lines at once}}. For instance, a 512\,GHz bandwidth would cover from $\lambda_{\rm rest}$\,$\sim$\,40\,$\mu$m to ``infinity'' in a source at $z=14$ ---that is, simultaneous coverage from Band 1 through 8 (the entire red line in the right panel), increasing the effective survey speed by a factor of $>$\,60, and ISM characterization by $N_{\rm lines}$ times. Finally, pushing the facility to its limits, ALMA2040 will be able to detect \oiii{}\,88\,$\mu$m in GS-z14-0-like progenitor galaxies up to $z\approx20$ (light blue line sensitivity and label annotation).}
    \label{fig:sensitivityplot}
\end{figure}

The large number of bright, massive galaxies at $z\,>\,10$, such as GLASS-z12/GHZ2 \citep{Naidu2022}, JADES-GS-z14-0 \citep{Carniani2024}, and MoM-z14 \citep{Naidu2026}, has challenged our understanding of early galaxy assembly, pointing to surprisingly vigorous star formation within the first few hundred million years after the Big Bang. Characterizing this process is now urgent, yet spatially resolving, or even simply detecting the cold gas that fuels star formation in the first galaxies remains an outstanding challenge ---one that line-of-sight $\mathrm{H\,I}$ absorption measurements in JWST spectra \citep{Heintz2025} cannot address. Traditional tracers of star formation fuel, like low-$J$ transitions of carbon monoxide (CO), shift to low radio frequencies at these redshifts, and are prohibitively faint and affected by the increased CMB temperature. Moreover, CO is less abundant in the metal-poor and dust-poor ISM of cosmic dawn galaxies (Section~\ref{s:metals}). The intense radiation fields produced by young stellar populations (Section~\ref{s:conditions}), combined with reduced dust shielding, allow ultraviolet photons to penetrate deeper into the ISM and dissociate the molecules. The \cii{}\,158\,$\mu$m line, as the dominant coolant of the cold ISM \citep{Wolfire2022}, arises from both the atomic and molecular phases, offering a powerful proxy for the cold reservoir ---atomic plus molecular--- that CO alone cannot capture at low metallicity \citep{Zanella2018, Heintz2021, DDS2025}. ALMA is the facility of choice for accessing this gas phase, yet at $z>8$ even \cii{} itself becomes difficult to detect (see gray lines in Figs.  \ref{fig:muvplot} and \ref{fig:sensitivityplot}).

Because of this, targeted ALMA campaigns have instead turned to the brighter, but purely ionized-gas tracer, \oiii{} 88$\mu$m line \citep{Harikane2020}, which has yielded detections in a handful of the most distant JWST-selected galaxies, including GHZ2 at $z=12.33$ \citep{Bakx2023, Zavala2024} and GS-z14-0 at $z=14.18$ \citep{Carniani2025, Schouws2025a}. The cooler gas probed by \cii{}, instead, remains out of systematic reach: detections struggle to advance beyond $z\approx7$--$8$ \citep[see][]{Carniani2020}, with MACS0416-Y1 still being the farthest \cii{}-detected galaxy to date \citep[$z=8.3$;][]{Bakx2020}, a result which is now six years old and predates JWST. 

An example of this limitation is the recent ALMA follow-up of GS-z14-0: given its star formation rate (SFR), local scaling relations \citep[][]{Delooze2014} predicted a \cii{} luminosity bright enough for an easy detection. Instead, a deep ALMA integration returned only an upper limit of $L_{\text{[C\,II]}} \approx 6\times10^{7}\ L_\odot$ \citep{Schouws2025b}. This suggests that \cii{} emission may be fainter than in lower-redshift sources with comparable SFRs. It is not yet clear whether this deficit reflects a rapidly exhausted cold gas reservoir, thus pointing towards a high star formation efficiency, or perhaps different metal abundance patterns and excitation conditions in the ISM \citep[e.g.][]{Casavecchia2025, Vallini2025}.

Whatever the reason, to gather an unbiased census of gas masses and gas fractions ($f_{\rm gas}=M_{\rm gas}/(M_{\rm gas} + M_{\rm *})$), and break the $z>10$ frontier, detecting \cii{} emission from representative, $L^\star$ galaxies at cosmic dawn will require a transformational leap in sensitivity. Existing constraints on $f_{\rm gas}$ up to $z\approx 7-8$ suggest that early galaxies were extraordinarily gas-rich. Results from \cii{} surveys like ALPINE \citep{Dessauges-Zavadsky2020} report $f_{\rm gas}$ rising steadily with redshift and plateauing at roughly $f_{\rm gas} \approx 0.6$ between $z = 4.4$ and $z = 5.9$. In the REBELS survey at $z \approx 7\text{--}8$, \citet{Aravena2024} find $f_{\rm gas} \approx 0.7$, while \citet{Fudamoto2026} recently exploited \oi{}\,145\,$\mu$m to constrain the gas fraction in four bright galaxies at $z \approx 7$, finding $f_{\rm gas}=0.3\text{--}0.8$, consistent with $[\text{C}\,\textsc{ii}]$-based estimates. Interestingly, large gas reservoirs are not limited to star-forming systems; the massive quiescent galaxy RUBIES-QG-z7 at $z = 7.3$ is characterized by $f_{\rm gas} > 0.2$, with much of the gas embedded within an extended halo despite its suppressed star formation \citep{Valentino2026}. Notably, \citet{Cescon2026} detected the CO(3-2) and CO(7--6) lines in the massive, dusty starburst REBELS-25 at $z = 7.3$. This detection translates into an exceptionally high $f_{\rm gas} = 0.95$. 
When pushing the redshift envelope to $z > 10$, measurements become highly uncertain, and gas fractions are derived from rough estimates of the dynamical mass. This is the case for GS-z14-0, for which only a loose constraint ($f_{\rm gas}<0.5$--$0.8$) is obtained from the poorly resolved width of \oiii{}\,88\,$\mu$m \citep{Carniani2025}, a line that, moreover, traces ionized rather than cold gas.

\begin{figure}[htbp]
    \centering
    \includegraphics[width=\textwidth]{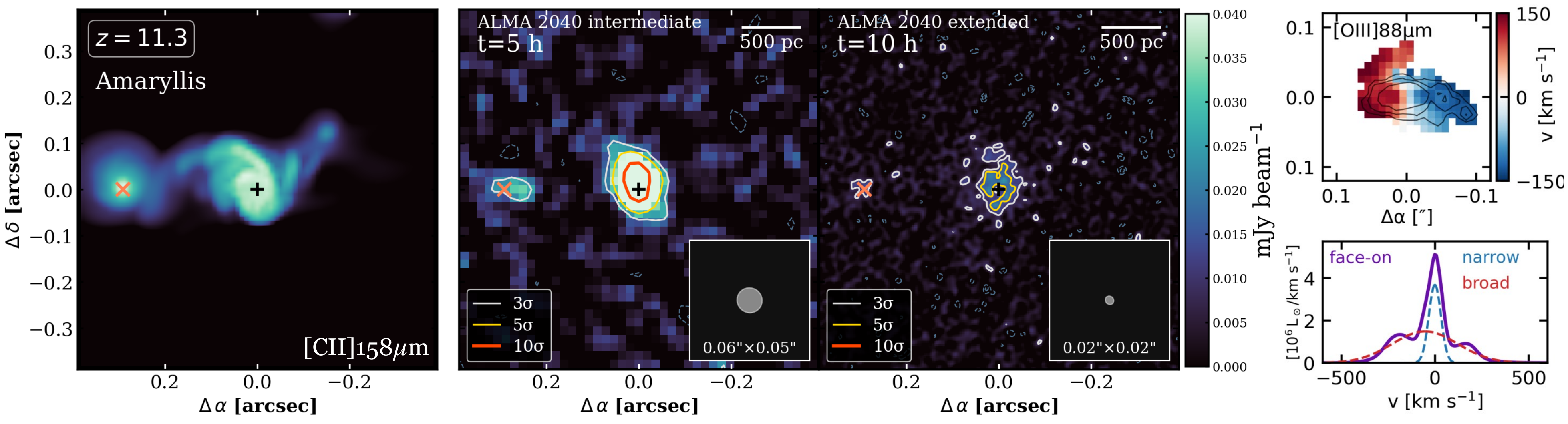}
    \caption{\textbf{Left:} [C\,\textsc{ii}] $158\,\mu\mathrm{m}$ emission of Amaryllis (\texttt{serra\_04:19:3635}) at $z=11.3$ \citep[][]{kohandel2025} extracted from the SERRA simulation \citep{Pallottini2022}. \textbf{Center:} GHOST simulation of an Amaryllis \cii{} observation with the `intermediate configuration' ($255\times12\,\mathrm{m}$ antennas, $12.5\,\mathrm{km}$ max baseline) and an integration time $t_{\text{int}}=5\,\mathrm{h}$. Solid contours indicate S/N$=3\sigma$ (white), $5\sigma$ (gold), and $10\sigma$ (red) levels. The synthesized beam is $0.06''\times0.05''\approx 235 \times 195\,\mathrm{pc}^2$. Amaryllis observed with the `extended configuration' ($28\,\mathrm{km}$ max baseline) and $t_{\text{int}}=10\,\mathrm{h}$ (synthesized beam $0.02''\times0.02''$ $\approx 80 \times 80\,\mathrm{pc}^2$). \textbf{Right:} The Amaryllis' \oiii{} velocity field shows an ordered $\pm150\,\mathrm{km\,s^{-1}}$ gradient across a compact gaseous disk. The face-on \oiii{} spectrum splits into a narrow component and a broad wing tracing an ionized outflow \citep[velocities relative to the narrow component;][]{kohandel2025}.}
    \label{fig:amaryllis_ghost_sims_v2}
\end{figure}

Yet, \oiii{}\,88\,$\mu$m is the brightest FIR line in the high-$z$ Universe and will remain the benchmark for the spectroscopic identification of $z>10$ sources detected across current and future facilities up to the 2040s. Resolving the line kinematics would directly alleviate the above limitation: the simulated \oiii{} profile and velocity field of Amaryllis (Fig.~\ref{fig:amaryllis_ghost_sims_v2}), whose \oiii{} luminosity is comparable to that of GS-z14-0 and GHZ2, show a velocity gradient from which a dynamical mass could be reliably derived at the angular resolutions achieved by ALMA2040.

Particularly, Fig.~\ref{fig:sensitivityplot} shows that the increase of at least 10$\times$ in point-source line sensitivity of ALMA2040 would potentially allow the detection of \oiii{}\,88\,$\mu$m in GS-z14-0-like progenitors up to $z=20$ (blue annotation), well beyond the current JWST record. Furthermore, by increasing the instantaneous bandwidth to at least 512\,GHz, ALMA2040 will have the ability to detect \textit{\textbf{at once}} all FIR emission lines covered from Bands 1 through 8 for sources like GS-z14-0 in under 25\,hours (see Fig.~\ref{fig:sensitivityplot}, red line), effectively becoming the ultimate spectroscopic-redshift identification and ISM characterization machine. The instantaneous multi-line coverage will allow stacking analyses to push to even deeper limits, enabling ISM studies of fainter source populations and galaxies beyond the redshift limits of individual detections. To summarize, our understanding of the gas content in high-$z$ galaxies rests on a heterogeneous mix of methods and biased samples ---ionized gas phase measurements via \oiii{}\,88\,$\mu$m, line-luminosity-to-cold-gas-mass scaling relations calibrated on \cii{}, dynamical mass estimates from scarcely resolved line widths, and, in a few cases, CO detections--- each carrying its own systematics.
Breaking this impasse requires moving beyond case-study targets toward a homogeneous census of gas masses and ISM physics in representative samples. A next-generation facility like ALMA2040 will be transformative in this respect, enabling multi-line follow-up of hundreds of cosmic dawn galaxies selected by current and future state-of-the-art facilities, and the discovery of new sources out to $z\approx 20$.

\subsubsection{Resolving the cold gas in \texorpdfstring{$z>10$}{z>10} galaxies down to 100 pc scales}
\label{s:resolving}

Knowing how much gas is available is only half the picture: an equally fundamental question is how efficiently primeval galaxies convert this gas into stars. 
One possibility is that in the ultra-dense, low-metallicity primordial ISM, star formation could have been remarkably rapid, with gas potentially cooling on extremely short time-scales \citep{Maio2010, Dekel2023}. Another hypothesis is that such powerful starbursts were able to expel gas and dust via radiative outflows, thus clearing star-forming regions and allowing them to appear very bright in the UV \citep{Ferrara2024}.
To quantify the star-formation efficiency in the early Universe, $\mathrm{SFE}=SFR/M_{\rm gas}$, or its inverse, the gas depletion time scales, $t_{\rm dep}$, we need a solid tracer of $M_{\rm gas}$. Current sub-mm capabilities allow the determination of $t_{\rm dep}$ in samples of galaxies only up to $z\approx 7$ \citep{Aravena2024}, or thanks to gravitational lensing \citep{Narita2026}, finding $t_{\rm dep}\approx$ 1 Gyr, broadly in agreement with theoretical predictions \citep{Maio2026}. Nevertheless, the scatter is large, with sources caught during bursts of star formation showing $t_{\rm dep}\approx 100$ Myr.


Another important aspect is that the global $t_{\rm dep}$ does not reveal the spatial distribution of star formation, nor how efficiently gas is converted into stars on small scales. This requires resolving the Kennicutt-Schmidt relation, which links $\Sigma_{\rm SFR}$ to $\Sigma_{\rm gas}$, and whether it evolves at cosmic dawn. Recent rest-frame UV and optical imaging from JWST already resolves the stellar side of this picture, often revealing highly compact, clumpy stellar cores in the first few hundred million years of cosmic time \citep{Adamo2025}. What is still missing is the matching high-resolution view of the gas. ALMA2040's ability to resolve morphologies and kinematics will be key to connecting gas turbulence to the efficiency of early star formation, thus moving beyond simple merger-versus-disk classifications \citep[][]{Rowland2024}, and to determine whether gas is smoothly distributed in these systems or clustered into self-gravitating, highly star-forming pockets.
Programs such as ALPINE-CRISTAL(-JWST) \citep{Herrera-Camus2025, Faisst2026} and REBELS-IFU  (Stefanon et al. in prep.), or those targeting lensed systems \citep[][]{Fujimoto2025}, represent the state-of-the-art. They have obtained \emph{kpc-scale} mapping of cold gas and star formation via joint ALMA \cii{} and JWST nebular-emission observations of massive, UV-bright galaxies at the end tail of the EoR. 
Expanding this kind of analysis at cosmic dawn requires a transformational facility, offering at least a $10\times$ improvement in sensitivity and at least a factor of a few in angular resolution over ALMA. The effect of such a transformational leap is shown in Fig. \ref{fig:amaryllis_ghost_sims_v2}. There we consider the $z=11.3$ simulated galaxy \emph{Amaryllis} extracted from the SERRA cosmological zoom-in simulation \citep{Pallottini2022}. \textit{Amaryllis} is a representative cosmic-dawn galaxy: with $M_\star = 8.8\times10^{8}\,\mathrm{M_\odot}$, $\mathrm{SFR}_{10\,\mathrm{Myr}} = 18\,\mathrm{M_\odot\,yr^{-1}}$, a mass-weighted stellar age of $\simeq 43$\,Myr, and $M_{\rm gas} = 4.5\times10^{8}\,\mathrm{M_\odot}$, it lies in the same regime as GHZ2 and GS-z14-0 \citep{kohandel2025}. As shown in Fig. \ref{fig:amaryllis_ghost_sims_v2} the ALMA2040 inter\-me\-dia\-te configuration will be able to spatially resolve both the galaxy and its satellite in only 5\,hours. Moreover, it will take ALMA2040 only 10\,h to catch hints of "spiral" features in the main source, down to 80\,pc scales, and also detect \cii{} from the companion. In the GHOST simulated extended configuration (28 km), the resolution is $\simeq80$\,pc at $z=11.3$, but in case of gravitational lensing, magnification would improve both the tangential resolution and the luminosity limit. That is, for lensed systems, ALMA2040 will achieve unprecedented cold gas measurements down to the tens-of-pc scales of stellar clumps at cosmic dawn (e.g. Cosmic Gems $z=9.3$ \citealt[][]{Messa2026}; $z\approx 11-12$ Misty Moons \citealt[][]{Nakane2025}).

\subsection{What were the ISM conditions of typical galaxies at cosmic dawn?}
\label{s:conditions}

At cosmic dawn, the physical and chemical conditions of the ISM undergo rapid evolution driven by the initial metal enrichment produced by the first starbursts. Characterizing the fundamental properties of the primordial ISM ---specifically the volume density ($n_{\rm H}$), kinetic temperature ($T_{\rm kin}$), metallicity, and the hardness and intensity of the radiation field ($U_{\rm ion}$)--- is key for understanding early galaxy assembly \citep{DDS2025}. Indeed, the translation of \cii{} luminosity into total cold gas mass relies on a conversion factor that is highly dependent on most of these properties \citep{Vallini2025, Casavecchia2025}. Unresolved detections in very compact sources blend physical regions together, washing out local variations and leading to estimates with large uncertainties. FIR and sub-mm fine-structure lines (Fig.~\ref{fig:sensitivityplot}) provide us with direct, dust-extinction-free, thermodynamical diagnostics capable of fully characterizing the multi-phase ISM properties of galaxies. The critical densities of several FIR lines are significantly lower than those of prominent rest-frame UV/optical lines targeted by JWST, thereby tracing more diffuse gas and enabling a complementary view of the ionized gas phase that is difficult or impossible to isolate at shorter wavelengths.

\begin{wrapfigure}[19]{r}{0.58\textwidth}
  \vspace{-2mm}
  \centering
  \includegraphics[width=\linewidth]{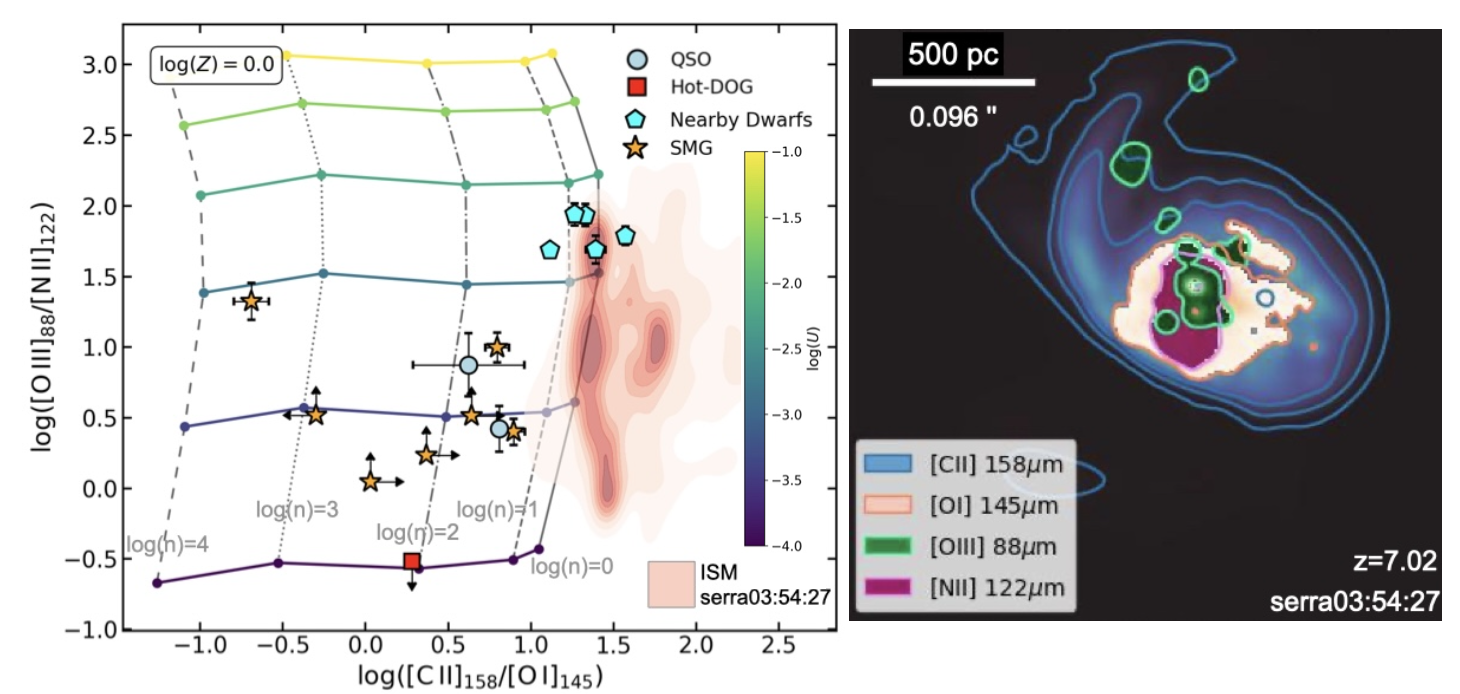}
  \caption{ALMA2040 will push studies of the first galaxies from individual line detections to spatially resolved, multi-line characterizations of their physical conditions. \textbf{Left:} \oiii{}$_{88}$/\nii{}$_{122}$ vs. \cii{}$_{158}$/\oi{}$_{145}$: photoionization models at $Z=Z_{\odot}$ (density dashed, ionization parameter color-coded) and ISM pixels (shaded) of the simulated \texttt{serra03:54:27} at $z=7.02$, with observed sources \citep{DDS2025}. \textbf{Right:} simulated line maps of the same galaxy \citep{Pallottini2022}.}
  \label{fig:ism}
  \vspace{-3mm}
\end{wrapfigure}

In the early Universe, current facilities only sample the most-luminous tail of the line-emitting population, limiting our understanding of this critical phase of structure formation. For instance, low \oiii{}-to-\cii{} line ratios reveal a population of evolved, steady star-forming galaxies at $z\,\sim\,7$ \citep{Algera2024}, standing in sharp contrast to the coeval population of lower-mass systems characterized by strong ionization fields and high gas densities \citep{Harikane2020}. To break the degeneracy among the multi-phase gas properties in high-$z$ galaxies and characterize their stellar populations, a next-generation facility capable of providing a broad suite of atomic and ionized emission lines is needed. For example, measuring two fine-structure transitions of the same species, such as \oiii{}\,52 and 88\,$\mu$m or \nii{}\,122 and 205\,$\mu$m, provides a metallicity independent measurement of the gas density \citep{Zavala2024, Parente2025}. Complementarily, the \oiii{}/\nii{} and \cii{}/\oi{} ratios jointly trace the ionization parameter and the gas density of PDRs, as shown by the diagnostic diagram in Figure~\ref{fig:ism}. Leveraging these diagnostics in EoR sources can be revolutionary, but only with sufficient spatial resolution. Simulations predict different emission morphologies depending on the gas tracer; hence, the position where the ISM regions fall in the diagram is hardly interpretable using a single-zone model. ALMA2040's superb spatial resolution, sensitivity, and instantaneous frequency coverage will allow us to capture and resolve all the key lines \emph{at once} (Fig.~\ref{fig:sensitivityplot}).

\subsection{How fast was the metal enrichment and molecular build-up in the early Universe?}
\label{s:metals}

The blue colors of $z>10$ galaxies suggest low dust attenuation, possibly due to small dust column densities or unusual dust extinction curves. Beyond dust, the time-scales for the formation of the first metals, forged through stellar nucleosynthesis perhaps in Pop-III stars, remain a key open question in our understanding of the early stages of chemical enrichment \citep{Maio2010}. 
While ALMA has already set upper limits on dust-to-stellar mass ratios of $< 2 \times 10^{-3}$ in the most distant galaxies, consistent with supernova dust yields y$_d < 0.3$ M$_{\odot}$ \citep{Algera2024}, much deeper, multi-band surveys are needed to establish the precise metal enrichment as well as dust formation and growth pathways. The transformational impact of ALMA 2040 for the study of dust at cosmic dawn is discussed in the \emph{\underline{Algera et al.}} white paper. Early observations with JWST have also started the characterization of the mass-metallicity relation at high-$z$, highlighting how pristine gas inflows and early stellar yields govern primordial systems \citep{Sarkar2025}. However, measuring absolute $Z_{\rm gas}$ at the highest redshifts remains challenging due to the physical degeneracies between gas density, temperature, and ionization field intensity. Furthermore, rest-optical line diagnostics are biased towards the least dusty {H\,{\sc ii} regions. A facility capable of capturing the brightest FIR fine-structure transitions, unaffected by dust extinction and also coming from colder gas, will be essential to eliminate these biases and degeneracies. Combined with rest-optical lines, FIR line diagnostics will help reveal whether early gas enrichment was set by core-collapse supernovae or by top-heavy Pop~III stars. Differential elemental abundances can also be used to decode how chemical enrichment proceeded, serving as a cosmic clock for early nucleosynthesis. For instance, JWST spectroscopy has uncovered an unexpected population of nitrogen and carbon enhanced galaxies at redshifts as high as $z\,\sim\,12$, showcasing extreme N/O ratios and elevated C/O at sub-solar metallicities \cite[][]{Kobayashi2024, NC2025, Zhu2026}. While extreme N-enrichment is in some cases attributed to clustered star formation (sometimes accompanied by AGN signatures), or specific high-mass-loss stellar phases like Wolf-Rayet stars, elevated C/O ratios carry the unique signature of chemical enrichment by Pop~III or metal-poor massive stars \citep{Hsiao2025}. Extending this cosmic clock to the earliest cosmic times to identify the very first generations of starbursts requires a transformational leap in sub-mm sensitivity to detect emission lines such as \oiii{}\,52 and \niii{}\,57\,$\mu$m (Fig.~\ref{fig:sensitivityplot}), that can directly trace secondary enrichment tracks like the C/O and N/O abundance while breaking electron density degeneracies \citep[e.g.,][]{Chartab2022}. Theoretical models and observations at high redshift suggest a complex interplay where dust content depends on localized gas density and efficient dust growth through coagulation, competing against shattering and destruction from supernova shock waves \citep{Popping2017, Aoyama2020, Narayanan2026}. Indeed, the apparent lack of dust continuum in several ultra-early systems may reflect efficient dust clearing driven by strong radiation pressure from highly efficient starbursts or inefficient grain growth \citep{Schouws2025a, Algera2026}. The dust-to-metal ratio, and its connection to absolute elemental abundances, can also serve as a key independent diagnostic of early supernova energetics and progenitor type.
Indeed, simulations show that different Pop~III progenitors ---standard core-collapse, massive core-collapse, and pair-instability supernovae--- produce unique abundance ratios ([C/Fe], [Si/C], and [Si/O]), providing a key indirect diagnostic to identify the stellar sources enriching high-redshift systems \citep{Ma2017}. In summary, by mapping the co-spatial distribution of dust continuum (see  \emph{\underline{Algera et al.}}) and absolute metal column densities at sub-kpc scales, with ALMA2040 we will be able to quantify early chemical enrichment, providing the necessary observational constraints to characterize the first generation of stellar explosions.

\subsection{What is the role of feedback on the formation of the first galaxies?}
\label{s:feedback}

During the first few hundred million years of cosmic time, the shallow gravitational potentials of the earliest dark matter halos made primordial galaxies exceptionally fragile to the competing forces of baryonic accretion and energetic feedback. The rapid formation of the first stars and the growth of supermassive black hole seeds injected massive amounts of radiative and mechanical energy into the pristine ISM \citep{Napolitano2025}. These feedback processes operate as primary regulators in the early baryon cycle, driving powerful galactic outflows that pierced through dense, star-forming cores, counteracting the continuous inflow of cold filamentary streams from the circumgalactic medium (CGM). 
Uncovering the detailed pathways of the feedback processes relies entirely on resolving the gas kinematics at high spatial and spectral resolution. As discussed in \underline{\emph{Kaasinen et al.}}, spatially resolved velocity fields make it possible to separate non-circular motions from turbulent broadening, and hence to test whether stellar feedback or gravitational instabilities dominate the turbulent energy budget. Signatures of ionized outflows have been found through rest-frame optical H$\alpha$ and \oiii{} in galaxies up to $z\,\sim\,9$ \citep{Xu2025}, and extreme AGN signatures in systems at $z\,>\,10$ \citep{Maiolino2024}. However, tracing the colder, neutral component of these outflows requires high-resolution kinematics of FIR fine-structure lines. While the cores of the \cii{} and \oiii{} lines seem to indicate ordered rotation and/or turbulent dispersion, the presence of faint, broad line components with $\Delta v>500$\,km\,s$^{-1}$ provides strong evidence of outflowing gas \citep{Algera2026}. In the simulated \oiii{}\,88\,$\mu$m profile of Amaryllis (Fig.~\ref{fig:amaryllis_ghost_sims_v2}), a narrow core (FWHM\,$\simeq74$\,km\,s$^{-1}$) tracing the ordered motions of the galaxy coexists with a broad component (FWHM\,$\simeq410$\,km\,s$^{-1}$): the sensitivity and spectral coverage of ALMA2040 will make it possible to separate the two and show whether the broad component is associated with an outflow;
also, it will yield $\approx$3 resolution elements across the 350 pc diameter of the \oiii{} disk.

Determining the efficiency of early feedback processes, particularly in low-mass systems, remains an unattained goal even nowadays ---one that cannot be answered unless we place robust constraints on critical quantities such as mass loading factors as a function of the galaxy mass. Powerful feedback events dynamically reshape the multi-phase ISM relentlessly by carving out highly ionized cavities and driving gas into the outer boundaries of the host galaxy. As this metal-enriched gas material is pushed outwards, a fraction of it will likely fall back through fountain-like streams, as observed in local galaxies \citep[e.g.,][]{Fraternali2017}, and the rest will populate the CGM with diffuse, extended halos, directly influencing the cooling efficiency and thermal state of the subsequent accretion episodes (for details see the ALMA2040 white paper on CGM, \underline{\emph{Butler et al.}}). Indeed, the extended \cii{} halos already detected around early galaxies \citep{Ginolfi2020} could be signatures of extraplanar gas ejected by stellar feedback \citep{Pizzati2020} or of unresolved star-forming satellite galaxies. By spatially resolving the kinematics of different gas phases, we can measure the mass-loading factors and physical reach of early galactic winds, providing a direct handle on how the Universe's first feedback loops coupled to the ISM and CGM. 

\section{ALMA2040: Technical Requirements}
\subsection{Simulations with GHOST}
\label{s:ghost}

To quantify the capabilities of ALMA2040, we used the Generalised Hybrid Observation Support Tool \citep[GHOST][]{}\footnote{\url{https://alma-ghost.jb.man.ac.uk}}. As input sky model, we used the synthetic \cii{} 158 $\mu$m map of Amaryllis a prototypical digital twin of cosmic dawn galaxies \citep{kohandel2025} with $L_{\rm [CII]} = 5.4 \times 10^{7} \, L_{\odot}$ and a \cii{} half light radius of 209 pc (Fig.~\ref{fig:amaryllis_ghost_sims_v2}, left). The system comprises a main galaxy and a satellite, separated by $\sim$0.27'' ($\sim$1.1 kpc in projection). We simulated two potential ALMA2040 configurations observing at 154.4\,GHz (the frequency of redshifted \cii{}) over a bandwidth of 0.19 GHz ($\sim 370\rm \,km\,s^{-1}$, matched to the source's intrinsic velocity width). The \emph{intermediate} configuration, imaged with Briggs weighting (robust $=-0.25$, favoring resolution) and run for just 5\,hours on source, already delivers an impressive result: a synthesized beam of $0.055'' \times 0.054''$ ($\sim215\times209$ pc at $z=11.3$) and a map rms of 7.2$\, \mu \rm Jy\,beam^{-1}$, robustly detecting the main galaxy at $\sim 13\sigma$ peak significance, with the satellite present at $\sim4.5 \sigma$ (Fig.~\ref{fig:amaryllis_ghost_sims_v2}, central panel). In only 5\,hours, ALMA2040 both detects and begins to spatially resolve the merging system ---a regime that, as discussed in Section~\ref{s:content}, remains inaccessible to current ALMA. Pushing further, the \emph{extended} configuration, imaged with Briggs weighting (robust$=1$, favoring sensitivity) and run for 10\,hours on source, reaches a beam of $0.019'' \times 0.017''$ ($\sim76 \times68$ pc) ---to our knowledge the first demonstration that ALMA2040 can spatially resolve a galaxy on sub-100 pc scales while still detecting it at $z=11.3$ (Fig.~\ref{fig:amaryllis_ghost_sims_v2}, right panel). The main galaxy remains detected at $\sim 6.7 \sigma$ and the satellite at $\sim 4 \sigma$, despite the finer beam spreading the same flux over more independent resolution elements. 

\section{Synergies with current and future facilities}
\label{s:synergies}
 
ALMA2040 will transform our view of the ISM in early galaxies at cosmic dawn, and will maximize its scientific return through powerful synergies with current and future facilities, probing gas phases across different wavelengths, epochs, and physical scales. Here we discuss those with the ELT, PRIMA, and ngVLA. ELT/HARMONI will probe rest-UV emission lines from early galaxies, enabling spatially resolved spectroscopy and a comparison between their ionized phase and the neutral/molecular gas mapped by ALMA2040. The recently approved IR NASA mission PRIMA will be uniquely suited for synergies with ALMA2040. PRIMA's deep surveys will deliver rest-frame MIR spectra for thousands of galaxies at cosmic noon ---galaxies that will need to be followed up with FIR observations to study their cold gas content and properties. Together, PRIMA and ALMA2040 will track ISM evolution and metal enrichment across cosmic time using the same set of line tracers at different epochs. Finally, the ngVLA will target low-$J$ CO transitions in EoR galaxies that will complement the high-$J$ CO transitions observed by ALMA2040. 

\section*{Acknowledgements}
This work was developed as part of the broader ALMA2040 community effort. We acknowledge the Lorentz Center and NOVA for supporting the community workshops and coordination activities that helped initiate and shape the ALMA2040 White Paper process. The authors would like to thank the internal and external reviewers for their valuable input.\\

\bibliographystyle{apj_mod}
{\bibliography{bibliography}}

\subsection*{Affiliations}

{
\footnotesize{
\noindent $^{1}$ INAF, Osservatorio di Astrofisica e Scienza dello Spazio, Via P. Gobetti, 93/3, 40129 Bologna, Italy \\
\noindent $^{2}$ Foundation for Research and Technology--Hellas (FORTH), Heraklion, GR-70013, Greece \\
\noindent $^{3}$ SKA Observatory, Jodrell Bank, SK11 9FT, UK \\
\noindent $^{4}$ Institute of Astronomy and Astrophysics, Academia Sinica, 11F of Astronomy-Mathematics Building, Roosevelt Rd, Taipei 106319, Taiwan, R.O.C. \\
\noindent $^{5}$ Instituto de Estudios Astrofisicos, Universidad Diego Portales, Santiago, Chile \\
\noindent $^{6}$ DARK, Niels Bohr Institute, University of Copenhagen, Jagtvej 155, 2200 Copenhagen, Denmark \\
\noindent $^{7}$ European Southern Observatory (ESO), Karl-Schwarzschild Str. 2, 85748 Garching, Germany \\
\noindent $^{8}$ Department of Physics and Astronomy, Chalmers University of Technology, 412 96 Gothenburg, Sweden \\
\noindent $^{9}$ University of Ljubljana, Faculty of Mathematics and Physics, Jadranska ulica 19, SI-1000 Ljubljana, Slovenia \\
\noindent $^{10}$ Kapteyn Astronomical Institute, University of Groningen, P.O. Box 800, 9700AV Groningen, The Netherlands \\
\noindent $^{11}$ Scuola Normale Superiore, Piazza dei Cavalieri 7, 56126 Pisa, Italy \\
\noindent $^{12}$ Leiden Observatory, Leiden University, P.O. Box 9513, 2300 RA Leiden, The Netherlands \\
\noindent $^{13}$ Observatoire de Paris, LUX, Collège de France, CNRS, PSL University, Sorbonne University, 75014, Paris, France \\
\noindent $^{14}$ Instituto de F\'{i}sica y Astronom\'{i}a, Universidad de Valpara\'{i}so, Gran Breta\~{n}a 1111, Valpara\'{i}so 2360102, Chile \\
\noindent $^{15}$ Sorbonne Université, UPMC Paris 6 and CNRS, UMR 7095, Institut d’Astrophysique de Paris, 98b bd. Arago, 75014 Paris, France \\
\noindent $^{16}$ International Centre for Radio Astronomy Research, University of Western Australia, ICRAR M468, 35 Stirling Hwy, Crawley 6009, Western Australia, Australia \\
\noindent $^{17}$ Department of Astronomy, University of Geneva, Chemin Pegasi 51, 1290 Versoix, Switzerland \\
\noindent $^{18}$ IPAC, California Institute of Technology, 1200 E. California Boulevard Pasadena, CA 91125, USA \\
\noindent $^{19}$ Centro de Astrobiología (CAB), CSIC-INTA, Carretera de Ajalvir km 4, Torrejón de Ardoz, E-28850, Madrid, Spain \\
\noindent $^{20}$ Université de Strasbourg, CNRS, Observatoire astronomique de Strasbourg, UMR 7550, F-67000 Strasbourg, France \\
\noindent $^{21}$ Center for Frontier Science, Chiba University, 1-33 Yayoi-cho, Inage-ku, Chiba 263-8522, Japan \\
\noindent $^{23}$ Dunlap Institute for Astronomy and Astrophysics, 50 St. George Street, Toronto, Ontario, M5S 3H4, Canada \\
\noindent $^{24}$ Division of Physics, Faculty of Pure and Applied Sciences, University of Tsukuba, Tsukuba, Ibaraki 305-8571, Japan \\
\noindent $^{25}$ Waseda Research Institute for Science and Engineering, 3-4-1 Okubo, Shinjuku, Tokyo 169-8555, Japan \\
\noindent $^{26}$ DTU Space, Technical University of Denmark, Elektrovej 327, DK2800 Kgs. Lyngby, Denmark \\
\noindent $^{27}$ Cosmic Dawn Center (DAWN), Denmark \\
\noindent $^{28}$ Niels Bohr Institute, University of Copenhagen, Jagtvej 128, 2200 Copenhagen N, Denmark \\
\noindent $^{29}$ Kavli Institute for Cosmology, University of Cambridge, Madingley Road, Cambridge CB3 0HA, UK \\
\noindent $^{30}$ Cavendish Laboratory, University of Cambridge, 19 J. J. Thomson Avenue, Cambridge CB3 0HE, UK \\
\noindent $^{31}$ Instituto de Astronomía y Ciencias Planetarias, Universidad de Atacama, Copayapu 485, Copiapó, Chile \\
\noindent $^{32}$ Telespazio UK for the ESA, ESAC, Camino Bajo del Castillo s/n, E-28692 Villanueva de la Ca\~nada, Madrid, Spain. \\
\noindent $^{33}$ Scuola Internazionale Superiore Studi Avanzati (SISSA), Physics Area, Via Bonomea 265, 34136 Trieste, Italy \\
\noindent $^{34}$ INAF, Osservatorio Astronomico di Trieste, Via G. Tiepolo 11, 34143 Trieste, Italy \\
\noindent $^{35}$ Institute for Fundamental Physics of the Universe (IFPU), Via Beirut 2, 34151 Trieste, Italy \\
\noindent $^{36}$ INAF—Istituto di Radioastronomia—Italian ALMA Regional Centre, Via Gobetti 101, 40129 Bologna, Italy \\
\noindent $^{37}$ Department of Astronomy, University of Michigan, 1085 South University Avenue, Ann Arbor, MI 48109, USA \\
\noindent $^{38}$ Dipartimento di Fisica “Enrico Fermi”, Università di Pisa, Largo Bruno Pontecorvo 3, Pisa I-56127, Italy \\
\noindent $^{39}$ Pole Scientific, University Catholic of Lyon- University of Lyon, 10 place des Archives 69002, Lyon, France \\
\noindent $^{40}$ Instituto de F\'isica Fundamental, CSIC, Calle Serrano 123, 28006 Madrid, Spain \\
\noindent $^{41}$ Dipartimento di Fisica e Astronomia, Università di Bologna, Via Gobetti 93/2, 40129 Bologna, Italy \\
\noindent $^{42}$ Institut f\"ur Astrophysik, Universit\"at zu K\"oln, Z\"ulpicher Straße 77, D-50937 K\"oln, Germany \\
\noindent $^{43}$ Cluster of Excellence “Our Dynamic Universe” (DYNAVERSE) \\
\noindent $^{44}$ Departament d’Astronomia i Astrofisica, Universitat de Valencia, C. Dr Moliner 50, E-46100 Burjassot Valencia, Spain \\
\noindent $^{45}$ Department of Astronomy, Columbia University, New York, NY 10027, USA \\
\noindent $^{46}$ Center for Computational Astrophysics, Flatiron Institute, 162 Fifth Avenue, New York, NY 10010, USA \\
\noindent $^{47}$ School of Physics and Astronomy, Lancaster University, Lancaster, LA1 4YB, UK \\
}
}

\end{document}